\documentclass[journal abbreviation]{copernicus}
\usepackage{amsmath}
\usepackage{aas_macros}

\begin{document}

\title{Modelling the steady state spectral energy distribution of the BL-Lac Object {PKS 2155-30.4} using a selfconsistent SSC model}

\author[1]{M.~Weidinger}
\author[1]{M.~R\"uger}
\author[1]{F.~Spanier}

\affil[1]{Institut f\"ur Theoretische Physik und Astrophysik\\Universit\"at W\"urzburg\\Am Hubland\\97074 W\"urzburg}

\correspondence{M.~Weidinger\\ mweidinger@astro.uni-wuerzburg.de}

\runningauthor{}
\runninghead{M. Weidinger et al.: Modelling the steady state SED of PKS 2155}

\received{6 April 2009}
\revised{19 October 2009}
\accepted{23 November 2009}
\published{13 January 2010}


\firstpage{1}

\maketitle

\begin{abstract}
In this paper we present a fully selfconsistent SSC model with particle acceleration due to shock and stochastic acceleration (Fermi-I and Fermi-II-Processes respectively) to model the quiescent spectral energy distribution (SED) observed from {PKS 2155}. The simultaneous August/September 2008 multiwavelength data of H.E.S.S., Fermi, RXTE/SWIFT and ATOM give new constraints to the high-energy peak in the SED concerning its curvature. We find that, in our model, a monoenergetic injection of electrons at $\gamma_0 = 910$ into the model region, which are accelerated by Fermi-I- and Fermi-II-processes while suffering synchrotron and inverse Compton losses, finally leads to the observed SED of {PKS 2155-30.4} shown in \citet{fermi09}. In contrast to other SSC models our parameters arise from the jet's microphysics and the spectrum is evolving selfconsistently from diffusion and acceleration. The $\gamma_0$-factor can be interpreted as two counterstreaming plasmas due to the motion of the blob at a bulk factor of $\Gamma = 58$ and opposed moving upstream electrons at moderate Lorentz factors with an average of $\gamma_u \approx 8$.
\end{abstract}


\introduction[INTRODUCTION]
Among the class of active galactic nuclei (AGN), blazars are showing a spectral energy distribution (SED) that is strongly dominated by nonthermal emission across a wide range of wavelengths, from radio waves to gamma rays, and rapid, large-amplitude variability. Presumably, these characteristics are due to a highly relativistic jet which covers a small angle to the line-of-sight, emitting the observable Doppler-boosted synchrotron and inverse Compton radiation.\\
In SSC models the characteristic double humped spectra of blazars are explained by electrons in the jet emitting synchrotron radiation while being accelerated in a magnetic field, which gives the first peak in the SED. These high energy electrons upscatter the very same synchrotron photons to $\text{TeV}$ energies due to the inverse Compton effect, resulting in a second peak in the SED. Another approach to the explanation of the double humped structure are proton initiated electromagnetic cascades \citep[e.g.,][]{mannheim93} or external Compton models.\\
The key issue is to understand which physical mechanisms are leading to such SEDs. In particular that means to explain the ultrarelativistic electron spectra within the jet, which are believed to be responsible for the gamma radiation.\\
The high peaked BL Lac objects (HBLs) as a subclass of blazars show a peak in their SED in the X-ray regime, suggesting that an inverse Compton peak should occur at correspondingly high gamma-ray energies. In fact, a large fraction of the known nearby HBLs have already been discovered with Cerenkov telescopes, such as H.E.S.S., MAGIC, and VERITAS. Since 2008 the Fermi satellite measures at these high gamma-ray energies. The energy range of the Fermi data is slightly different from the H.E.S.S. and VERITAS-Telescopes which gives new constraints to the SEDs.\\
The first Fermi data published is from PKS 2155-30.4, a HBL at redshift $z=0.117$ (luminosity distance: {$d_{\text{L}} = 1.67 \cdot 10^{27} ~ \text{cm}$}) \citep{fermi09}.\\ \\
We present a selfconsistent SSC model that is not only able to model the SED of PKS 2155-30.4 shown in \citet{fermi09} but also to partly explain the ``ad-hoc'' injected particle spectra of many SSC models. Therefore we introduce and solve the kinetic equation describing the synchrotron-self-Compton emission numerically in two different zones within the jet (see section \ref{model}). We use the exact Klein-Nishina cross section which is important at the relevant very high gamma-energies to describe the inverse Compton radiation and energy losses of the electrons. The emphasis lies on the accurate treatment of the two possible particle acceleration mechanisms (Fermi-I- and Fermi-II) which are able to produce high energy electrons as well as on the selfconsistent treatment of the radiation processes.

\section{THE MODEL}
\label{model}

\subsection{Model geometry}
We extend the well-established SSC model by Fermi-I and Fermi-II acceleration mechanisms to a selfconsistent SSC model with two zones in a nested setup. Both regions (the acceleration- and the radiation zone) forming the blob are assumed to be spherical and homogeneous containing isotropically distributed non-thermal electrons and a randomly oriented magnetic field. The acceleration zone is assumed to be spatially significantly smaller than the surrounding radiation zone. Furthermore every electron leaving the accelerationzone enters the radiation zone. These assumptions are common place in SSC models \citep[e.g.,][]{kirk98}.\\
To derive the kinetic equations describing the time evolution of $n_e(\gamma)$, $N_e(\gamma)$ ($n_e$ in the acceleration zone, $N_e$ in the radiation zone) as the differential electron densities we use the one dimensional diffusion approximation (eq. \eqref{diffusionapproximation}) of the relativistic Vlasov equation \citep[e.g.][]{schlickeiser02}, which is applicable due to the assumptions made above.
\begin{align}
 \label{diffusionapproximation}
 \frac{\partial}{\partial t} f(p, t) & = \frac{1}{p^2}\frac{\partial}{\partial p} \left[ F\left(p, f, \frac{\partial}{\partial p}f \right) \right] + S(p,t)~,
\end{align}
where $f(p, t)$ is a particle distribution function, with the particle's momentum value $p$. $F$ describes the contributing processes, such as synchrotron radiation or acceleration, in momentum space. Catastrophic particle gains and losses are considered via $S(p, t)$.\\
Making use of the relativistic approximation $E \approx pc = \gamma mc$ and the relation $n(p, t) = 4\pi p^2 f(p, t)$ one can derive the kinetic equations governing the model.
\subsection{Kinetic equations}
\subsubsection{Acceleration zone}
While the blob propagates through the jet, electrons are continuously injected into the acceleration zone when considering the blob's rest frame, leading to an injection function
\begin{align}
 \label{injection}
 Q_{\text{inj}}(\gamma , t) & := Q_0 \delta(\gamma - \gamma_0) \vartheta{(t-t_0)}~~,~t_0 = 0
\end{align}
which we assume monoenergetic and time independent. These low- to mid-energy electrons are accelerated systematically and stochastically due to Fermi-I and Fermi-II processes while suffering synchrotron and inverse Compton losses. Energy losses due to inverse Compton scattering are calculated using the full Klein-Nishina cross section, see eq. \eqref{kn2}. This leads to $P_{\text{IC}}(\gamma)$ given in eq. \eqref{iclosses} with the corresponding radiation field $n_{\text{PH}}$ of the acceleration zone. Due to the non equilibrium of magnetic and radiative energy in the acceleration zone the energy losses via inverse Compton scattering can become quite significant and must not be neglected, also the Thomson limit is not appropriate here. The synchrotron losses are calculated using eq. \eqref{synchrotronlosses} from \citet{ginzburg69} for isotropic particle distributions
\begin{align}
 \label{synchrotronlosses}
 P_s(\gamma) & = \frac{1}{6 \pi} \frac{\sigma_{\text{T}}B^2}{mc}\gamma^2 = \beta_s \gamma^2
\end{align}
with the Thomson cross section $\sigma_{\text{T}}$. According to \citet{schlickeiser84} particle acceleration via parallel shockfronts and stochastic acceleration caused by scattering at Alfv\'en waves leads to
\begin{align}
 \label{acceleration1}
  F\left(p, f, \frac{\partial}{\partial p}f \right) & = p^4 \frac{v_A^2}{9K_{||}} \frac{\partial f}{\partial p} +p^3 \frac{v_S^2}{4K_{||}} f
\end{align}
for the function $F$. With the parallel spatial diffusion coefficient $K_{||}$, which is momentum independent for hard spheres and the characteristic speeds $v_A$ for the Alfv\'en mediated stochastic acceleration and $v_S$ for parallel shockfronts. Substituting $p \rightarrow \gamma$ in eq. \eqref{diffusionapproximation} and eq. \eqref{acceleration1} according to the relativistic approximation mentioned above, one will finally find eq. \eqref{acczone1}; the kinetic equation of the acceleration zone.
\begin{align}
 \label{acczone1}
 \frac{\partial n_e(\gamma, t)}{\partial t} = & \frac{\partial}{\partial \gamma} \left[( \beta_s \gamma^2 + P_{\text{IC}}(\gamma) - t_{\text{acc}}^{-1}\gamma ) \cdot n_e(\gamma, t) \right] +\nonumber \\
& + \frac{\partial}{\partial \gamma} \left[ [(a+2)t_{\text{acc}}]^{-1}\gamma^2 \frac{\partial n_e(\gamma, t)}{\partial \gamma}\right] + \nonumber \\
& + Q_0(\gamma-\gamma_0) - t_{\text{esc}}^{-1}n_e(\gamma, t)~,
\end{align}
where the characteristic acceleration timescale $t_{\text{acc}}$ is given by
\begin{align}
 \label{acceleration2}
 t_{\text{acc}} & = \left(\frac{v_s^2}{4K_{||}}+2\frac{v_A^2}{9K_{||}} \right)^{-1}~.
\end{align}
Eq. \eqref{acceleration2} for $t_{\text{acc}}$ is a direct consequence of the derivation of eq. \eqref{acczone1} out of eq. \eqref{diffusionapproximation} using eqn. \eqref{acceleration1} and \eqref{synchrotronlosses}. The expression in eq. \eqref{acceleration2} includes the analytical timescale for non-relativistic shock acceleration. According to \citet{bednarz96} and especially to \citet{ellison90} the acceleration timescale for parallel relativistic shock waves decreases approximately by a factor of $3$. We did not take into account this behavior for it is unclear how the analytical expression looks like in that case. Secondly we are omitting the energy dependency of $t_{\text{acc}}$ using hard spheres for the plasma instabilities anyway. This issue is irrelevant for the modelling (for we are setting numerical values for $t_{\text{acc}}$) and the type of energy spectrum (a powerlaw) produced by Fermi-I acceleration is identical in the non-relativistic and relativistic case \citep{sokolov}. But it has to be kept in mind for the interpretation of the results.\\
The parameter $a \approx v_s^2/v_A^2$ determines the ratio of shock to stochastic acceleration. $t_{\text{esc}} = \eta R_{\text{acc}}/c$ is the characteristic timescale for electrons escaping from the acceleration region, where $\eta$ is an empirical factor set to $\eta = 10$ and $R_{\text{acc}}$ the radius of the acceleration sphere. All escaping electrons enter the radiation zone downstream the jet. The seperation in two zones can firstly explain the injected electron spectra and secondly takes account of a much more confined shock region for Fermi-I acceleration will probably not occur in the whole blob region when considering physical sources.\\Our model can be compared with the model presented by \citet{kata}. The kinetic equation (eq. 3 in their paper) is almost similar to the kinetic equation in the acceleration zone eq. \ref{acczone1}. One major difference to our model is their sole use of stochastic acceleration. In fact their model is the limit of our model for $a \to 0$. Additionally they limit themselves to radiation in the acceleration zone, which is useful when not taking into account shock acceleration. Besides that there are number of minor differences regarding the exact treatment of inverse Compton losses and the derivation of escape rates.\\Due to the small spatial extent the acceleration zone does not contribute to the SED directly, i.e. $n_{\text{ph}}(\nu)$ is only calculated in order to determine the inverse Compton loss rate for the electrons in the acceleration zone.

\subsubsection{Radiation zone}
The electrons are not accelerated here. Thus the kinetic equation takes the simple form
\begin{align}
 \label{radzone1}
 \frac{\partial N_e(\gamma, t)}{\partial t}  = & \frac{\partial}{\partial \gamma}\left[\left(\beta_s \gamma^2 + P_{\text{IC}}(\gamma)\right) \cdot N_e(\gamma, t) \right] -\nonumber \\
&- \frac{N_e(\gamma, t)}{t_{\text{rad,esc}}} + \left(\frac{R_{\text{acc}}}{R_{\text{rad}}} \right)^3\frac{n_e(\gamma, t)}{t_{\text{esc}}}~\text{.}
\end{align}
Electrons in the radiation zone suffer synchrotron ($\propto \beta_s \gamma^2$) and inverse Compton losses (eq. \eqref{iclosses}), other energy losses are irrelevant in jetsystems of such low electron density \citep[see e.g.][]{boettcher02b}.
\begin{align}
 \label{iclosses}
 P_{\text{IC}}(\gamma) & = m^3c^7h \int_{0}^{\alpha_{max}}{d\alpha \alpha \int_0^{\infty}{d\alpha_1 N_{\text{ph}}(\alpha_1) \frac{dN(\gamma,\alpha_1)}{dtd\alpha}}}
\end{align}
The integrals in eq. \eqref{iclosses} are solved numerically using the full Klein-Nishina cross section for a single electron given in eq. \eqref{kn2}. The photon energies are rewritten in terms of the electrons rest mass, i.e. $h \nu = \alpha m c^2$ for the scattered photons and $h \nu = \alpha_1 m c^2$ for the target photons. The integration bounds of the outer integral in eq. \eqref{iclosses} are a direct consequence of the kinematics. The non-trivial dependency of $P_{\text{IC}}(\gamma)$ from $N_{\text{ph}}$ and thus of $N_e$ from eq. \eqref{iclosses} makes the numerical treatment of the kinetic equations inevitable leading to a time resolved model. The loss rates for the electron distribution of PKS 2155 in the steady state are shown in Fig. 1, which indicates that in our case the inverse Compton losses would be slightly overestimated in the often used Thomson limit for all Lorentzfactors $\gamma$ because of the dependency on the photon field $P_{\text{thom}} \propto \int{d\nu \nu N_{\text{ph}}} \gamma^2$ and its special shape due to the modified injection at $\gamma_0 = 910$. This would not be the case for low energetic injected electrons and the resulting powerlaw-like photon distribution. For high Lorentzfactors $\gamma$ however the deviation of the Thompson limit for the inverse Compton scattering to the real Klein-Nishina treatment becomes more significant in each case.
\begin{figure}[t]
 \vspace*{2mm}
 \begin{center}
 \includegraphics[width=8.3cm]{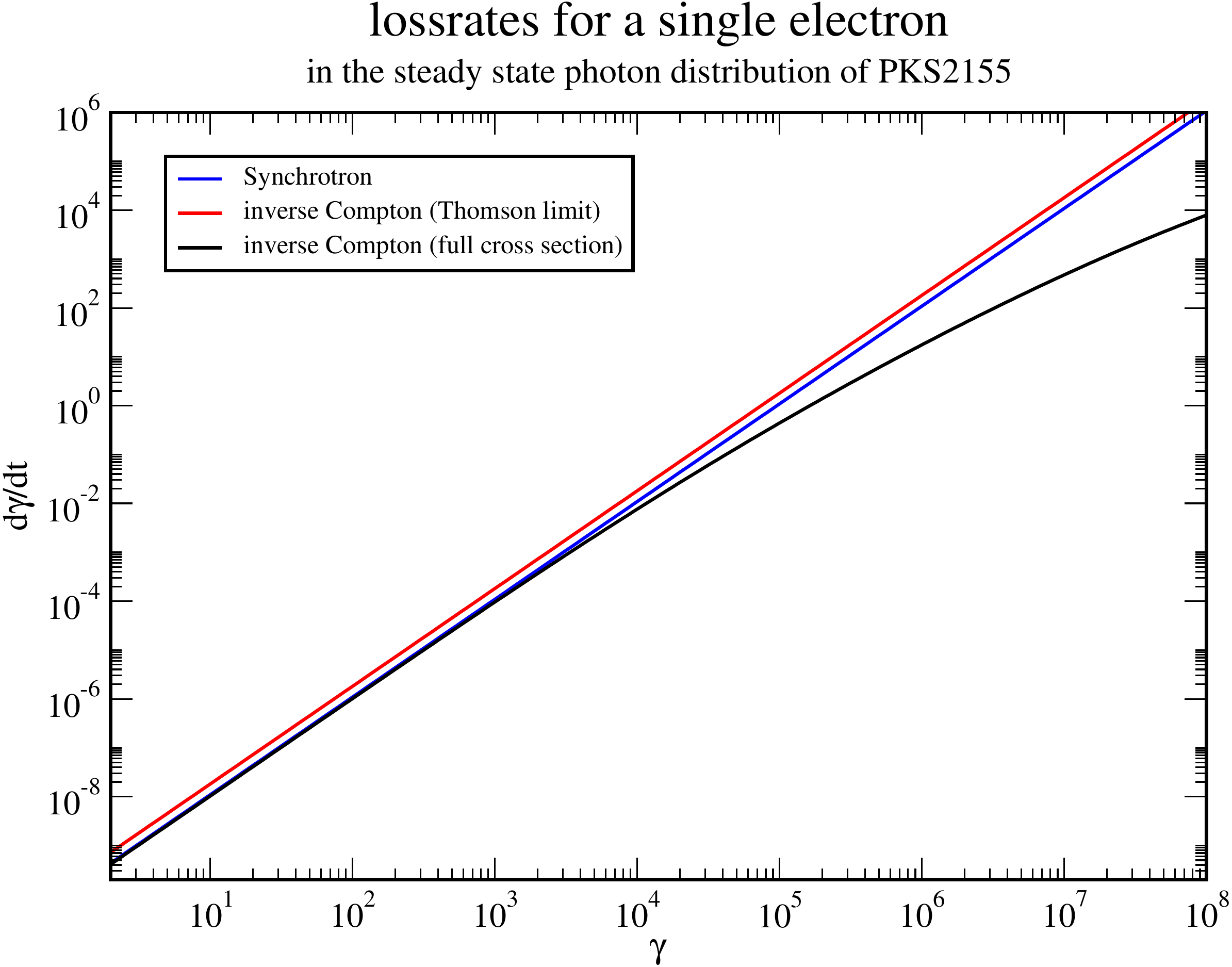}
 \end{center}
 \caption{Lossrates due to the inverse Compton effect for an electron of Lorentzfactor $\gamma$ in the model photon field of PKS 2155 (Thompson limit, red and used Klein-Nishina treatment, black) compared to the synchrotron losses, blue. Other losses like adiabatic cooling, pair production are irrelevant at typical SSC configurations.}
\end{figure}
Again electrons escaping the blob are taken into account via $t_{\text{esc,rad}} = \eta R_{\text{rad}}/c$ with the empirical factor $\eta$ set to $\eta = 10$. Both electronic PDEs are connected via the catastrophic particle loss/gain-term. The factor $\left(R_{\text{acc}}/R_{\text{rad}} \right)^3$ ensures particle conservation.\\ \\
To determine the time-dependent spectral energy distribution of blazars we solve the differential equation for the differential photon number density, obtained from the radiative transfer equation, including the corresponding terms with respect to the SSC model,
\begin{align}
 \label{radzone2}
 \frac{\partial N_{\text{ph}}(\nu, t)}{\partial t} & = R_s - c \alpha_{\nu} N_{\text{ph}}(\nu, t) + R_c - \frac{N_{\text{ph}}(\nu, t)}{t_{\text{ph,esc}}}~\text{.}
\end{align}
To describe the synchrotron photon production rate $R_s$ in a convenient way we use the well known Melrose approximation \citep[see e.g.,][]{pohl}
\begin{align}
 \label{melrose}
 R_s & = 1.8 \frac{\sqrt{3}e^3 B_{\bot}}{h \nu m c^2} \int{d\gamma N_e(\gamma, t) \left( \frac{\nu}{\nu_c(\gamma)}\right)^{\frac{1}{3}} e^{-\frac{\nu}{\nu_c(\gamma)}}}~.
\end{align}
with the characteristic frequency eq. \eqref{char} for an electron with a Lorentz factor of $\gamma$.
\begin{align}
 \label{char}
 \nu_c(\gamma) & = \frac{3\gamma^2 e B_{\bot}}{4\pi m c}
\end{align}
In optically thick regimes the emitted synchrotron radiation is absorbed by the emitting electrons itself. This is described by the synchrotron self absorption coefficient,
\begin{align}
 \label{SSA}
 \alpha_{\nu} & = \frac{1}{12} \frac{c}{\nu^2 e B}P_{\nu}(\gamma_c) \frac{N_e(\gamma_0, t)}{\gamma_0^2}\text{.}
\end{align}
(with $\gamma_c = f(\nu_c)^{-1}$). Here we made use of the monochromatic approximation \citep[e.g.,][]{fm66} for the synchrotron power:
\begin{align}
 P_{\nu}(\nu, \gamma) & = \frac{\sqrt{3} e^2 B_{\bot}}{mc^2} \frac{\nu}{\nu_c{\gamma}} \int_{0}^{\infty}{d\nu' K_{\frac{5}{3}}(\nu')}
\end{align}
\\In SSC models the second hump in the SED of a blazar is due to inverse Compton scattered photons by the synchrotron radiation emitting electrons themselves. Here the full Klein-Nishina cross section from \citet{bg70} is used to calculate the inverse Compton photon production rate. 
\begin{align}
\label{kn}
R_{c} =& \int d\gamma\, N_{e}(\gamma) \; \times \nonumber
\\ &\times \int d\alpha_1
\left[N_{\text{ph}}(\alpha_1)\frac{dN(\gamma,\alpha_1)}{dtd\alpha}- N_{\text{ph}}(\alpha)
\frac{dN(\gamma,\alpha)}{dtd\alpha_1}\right]
\end{align}
To fully exploit the Klein-Nishina cross section, eq. \eqref{kn}, we used the approximate inverse Compton spectrum of a single electron scattering off a unit density photon field \citep[e.g.,][]{jones68, jauch76}:
\begin{align}
\label{kn2}
\frac{dN(\gamma,\alpha_1)}{dtd\alpha} =& \frac{2 \pi r_0^2 c}{\alpha_1 \gamma^2} \bigg[ 2 q\ln q+(1+2q)(1-q)+\nonumber
\\&+\frac{1}{2} \frac{(4 \alpha_1 \gamma q)^2}{(1+4\alpha_1 \gamma
q)}(1-q)  \bigg]~,
\end{align}
with the electron's Lorentz radius $r_0 = e^2/(mc^2)$, the scattering parameter $q={\alpha}/(4\alpha_1 \gamma^2(1-\alpha/\gamma))$ and $0 \approx 1/(4\gamma^2)<q\leq 1$. Due to momentum and energy conservation this equation is valid for $\alpha_1<\alpha\leq 4\alpha_1\gamma^2/(1+4\alpha_1\gamma)$.\\
The last catastrophic term in eq. \eqref{radzone2} describes photons escaping from the emitting region, where
\begin{align}
t_{\text{ph, esc}}=\frac{3R_{\text{rad}}}{4c},
\end{align}
is the approximate escape time, with $R_{\text{rad}}$ the radius of the emitting blob. The escape time is chosen to be the light crossing time of the photons.\\
The photon lossrate due to the pair production of electrons and positrons is not taken into account for two reasons. Firstly it is insignificant compared to the dominating synchrotron and inverse Compton processes. This is a consequence of the relatively low density \citep{boettcher02b}. Secondly it would violate the selfconsistency of our model for positrons are not treated, hence violating energy conservation.
\\ \\
To compute the SEDs in our model we must shift the frame of reference from the blob to the observer. For a sphere of radius $R$ the observed flux at distance $r$ is
\begin{align}
 \label{obsflux}
 F_{\nu}^{obs}(r) & = \pi I_{\nu}^{obs} \frac{R^2}{r^2}~\text{.}
\end{align}
With the Lorentz boosted intensity $I_{\nu}^{obs} = \delta^3 I_{\nu}^{blob}$ due to the bulk motion with a doppler factor $\delta$ of the blob and the Lorentz transformed, red shifted frequency \mbox{$\nu^{obs} = \delta/(1+z) \nu$.} Where $I_{\nu}^{blob}$ is calculated from the photon unit density
\begin{align}
 I_{\nu}^{blob} & = \frac{h\nu c}{4\pi} N_{\text{ph}}(\nu)
\end{align}
for homogenous spheres.
\section{NUMERICS}
In our model we numerically solve the kinetic equations forward in time in order to obtain a model SED. The downstream motion of the electrons induces the sequence of solving the acceleration zone's equation before the kinetic equation of the radiation zone in each time step. The simple Euler scheme was found adequate to do the time integration.\\
In the acceleration zone we had to combine the Crank-Nicholson scheme \citep{press02} with Godunov's method to provide both correct treatment of the characteristics and stability for the derivation in $\gamma$. In the radiation zone the characteristic flows, due to the absence of acceleration, only in one direction making the Crank-Nicholson scheme sufficient.\\
With our carefully tested code it is possible to calculate the dynamics of SEDs in a range of 20 orders of magnitude. The implemented code complies particle conservation in each zone alone and both together as well as the conservation of the total energy (i.e. of the electrons and the photons) over typical simulation times with a maximum error of O(5\%). For negligible stochastic acceleration, i.e. $a \rightarrow \infty$, and without a radiation field, i.e. no inverse Compton losses, the steady state solution for the kinetic equation yields  
\begin{align*}
n_{\text{e,steady}}(\gamma) = C \frac{1}{\gamma^2} \left(\frac{1}{\gamma}-\beta_{\text{s}}t_{\text{acc}} \right)^{\frac{t_{\text{acc}}-t_{\text{esc}}}{t_{\text{esc}}}}
\end{align*}
with $\gamma_{\text{max}} = \left(t_{\text{acc}}\beta_{\text{s}}\right)^{-1}$ and the constant $C$ determined by the injection function $Q$. The implemented numeric model converges against this solution for sufficient simulation time. Additionally it was tested against the steady state analytical solution with Fermi-II processes given in \citet{schlickeiser84} with no significant deviations. Setting $t_{\text{esc,rad}} \rightarrow \infty$ and neglecting inverse Compton scattering the spectral index of the powerlaw part of the electron distribution in the radiation zone is (analytically) reduced by one compared to the one in the acceleration zone, which also was confirmed by the implemented code. The inverse Compton scattering rate was confirmed against the approximate analytical results (low energetic Thompson regime and extreme Klein-Nishina limit) before implementing. Concerning the photon distributions we validated the expected spectral indices in the steady state solution for the different frequency regimes, which together with the energy conservation between electrons and photons approves the integrity of the model.\\A detailed description of the used numeric techniques as well as the implemented model also in context with the variability of the sources will be given in a paper yet to be published.
\section{RESULTS}
The recent Fermi data give new constraints on the gamma-ray peak of the HBL PKS 2155-30.4 concerning its curvature. This is leading to a deep dip between the optical/X-ray and the gamma-ray peak. We are able to model the SED of PKS 2155-30.4 with our model by setting
\begin{align}
 \label{gamma0}
 \gamma_0 & = 910
\end{align}
for the monoenergetic injection into the acceleration zone. This is rather unusual but required to model the SED of PKS 2155. Such moderate but not small Lorentz factors can be explained e.g. by two counterstreaming plasmas. If the upstream electrons would be at rest, the bulk doppler factor of $\delta=116$ would automatically lead to $\gamma_0 = \Gamma \approx 58$. Assuming speculatively that the upstream electrons moving in the opposite direction of the blob with a mean velocity of $v_u$ hence a upstream Lorentz-Factor $\gamma_u$ the $\gamma_0$ factor in the blob's rest frame must be calculated according to the relativistic superposition:
\begin{align}
 \label{rel}
  \gamma_0 & = \sqrt{1-\left(\frac{\sqrt{\Gamma^2-1}\Gamma+\sqrt{\gamma_u^2-1}\gamma_u}{\Gamma^2+\gamma_u^2-1}\right)^2}^{-1}
\end{align}
Solving eq. \eqref{rel} for our setup we find $\gamma_u \approx 8$ for the upsteam electrons which are streaming towards the blob. The numerically solved steady state electron density in the acceleration zone is shown in Fig. 2. We also show the time development for a ``switched on'' injection, i.e. $\left.n_{\text{e}}(\gamma)\right|_{t<0} = 0~ \forall_{\gamma}$ and $Q=Q_0\delta(\gamma-\gamma_0)\vartheta(t)$, until the steady state is reached.
\begin{figure}[t]
 \vspace*{2mm}
 \begin{center}
 \includegraphics[width=8.3cm]{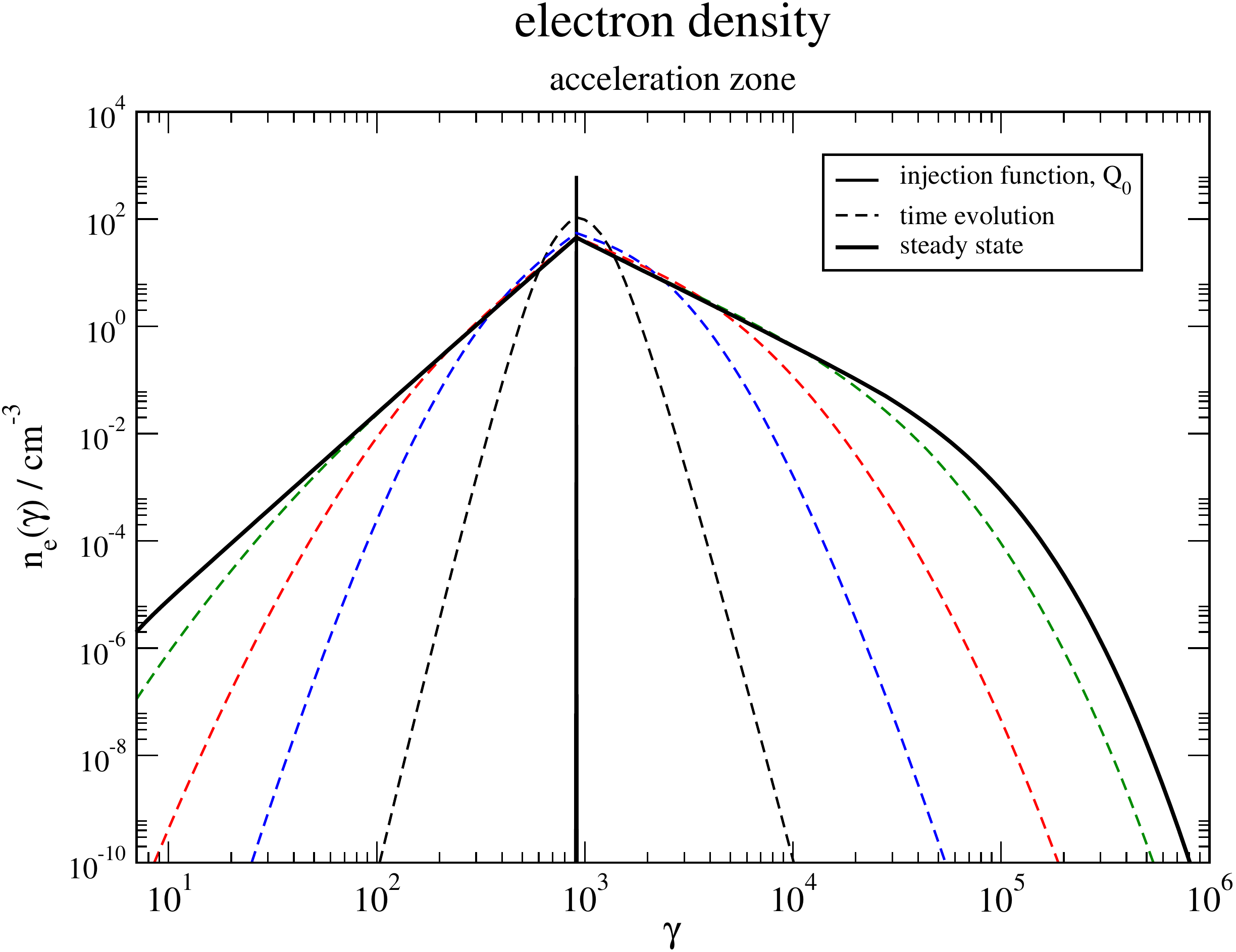}
 \end{center}
 \caption{Steady state electron distribution and its time development in the acceleration zone modelling the SED of PKS 2155-30.4 shown in Fig. 3 as arising from the injection function $Q_0$. The corresponding intrinsic times are $t=1000~\text{s}$ (dashed black), $t=5000~\text{s}$ (dashed blue), $t=1 \cdot 10^4~\text{s}$ (dashed red), $t=2 \cdot 10^4~\text{s}$ (dashed green). The steady state with its rising and falling powerlaw and the exponential cutoff at $\gamma \approx 10^5$ is reached at about $t=10^5~\text{s}$.}
\end{figure}
\\In Fig. 2 it can clearly be seen tvhat accelerating electrons using Fermi-I and Fermi-II processes leads to powerlaw electron distributions with an exponential cut-off often used as the ad-hoc injection function in onezone-SSC models \citep{boettcher02} (right side of $\gamma_0$ in Fig. 2), thus explaining them using the diffusion theory derived from plasma physics. By injecting electrons with eq. \eqref{gamma0} and significant stochastic acceleration (i.e. $a=O(1)$) we are also able to produce rising electron spectra before decreasing in a power-law and an exponential cut-off, like introduced in \citet{boettcher02b} (left side of $\gamma_0$ in Fig. 2). The Fermi-II processes are responsible for the rising power-law and exponential cut-off, whereas the ratio of $t_{\text{acc}}/t_{\text{esc}}$ determines the spectral index of the power-law at $\gamma>\gamma_0$. It can clearly be seen from Fig. 2 that the convergence against the steady state solution for the electron density begins relatively rapid while slowing down eventually. The simulation time when the steady state is reached corresponds to the escape time of the electrons in the acceleration zone. When concerning variability and time resolved lightcurves of blazars this is an advantage of the twozone model because the rising part in such lightcurves corresponds partially to the escape time of the acceleration zone (while the falling part is connected to the response time of the system, $t_{\text{rad,esc}}$).
\begin{table}[h]
 \caption{Chosen parameters for the model SED shown in Fig. 4 to fit the data \citep{fermi09} of PKS 2155-30.4.}
 \vskip4mm
 \centering
\begin{tabular*}{1\linewidth}{@{\extracolsep{\fill}}ccccccc}
\tophline
\tophline
$Q_0 (\text{cm}^{-3})$  & $B(\text{G})$ & $R_{\text{acc}}(\text{cm})$ & $R_{\text{rad}}(\text{cm})$ & $t_{\text{acc}}/t_{\text{esc}}$ & $a$ & $\Gamma$\\
\middlehline
\rule[-6pt]{0pt}{21pt}   $5.25 \cdot 10^{4}$ & $0.29$ & $3.0 \cdot 10^{13}$ & $6.3 \cdot 10^{14}$ & $1.55$ & $1$ & $58$\\
\bottomhline
\bottomhline
\end{tabular*}
\label{tab:2155}
\end{table}
\\An acceleration zone electron density as shown by the solid black curve in Fig. 2, leads to the desired broken power-law electron spectrum in the radiation zone which finally is able to model the SED of PKS 2155-30.4 (see Fig. 3 and Fig. 4). We used the parameters in Table \ref{tab:2155} for the model SED in Fig. 4 (black, solid line). The black dashed curve in Fig. 4 corresponds to a fit assuming a black body for the thermal contribution of the host galaxy thus the ATOM optical data is not to be taken into account for the SSC modelling. The curvature and deep dip in the model SED is a direct consequence of the rising part in the electron density of the acceleration zone. Thus it can be modeled by varying the ratio $a$ of shock to stochastic acceleration.  All the parameters in Table \ref{tab:2155} are consistent with the limits given via other observations and statistics, e.g. determination of $\Gamma$ using superluminal motion of Quasar jets.\\The recent Fermi, H.E.S.S. and ATOM data \citep{fermi09} have been averaged over a period of 14 days and show a lowstate of the HBL PKS 2155-30.4. This is confirmed by the \citet{multiwave02} data of H.E.S.S. a few years ago which show the same flux level as the recent data. We used the EBL studies described in \citet{primack05} to do the EBL deabsorption for the H.E.S.S. datapoints, a correction of the Fermi data is not necessary.
\begin{figure}[t]
 \vspace*{2mm}
 \begin{center}
 \includegraphics[width=8.3cm]{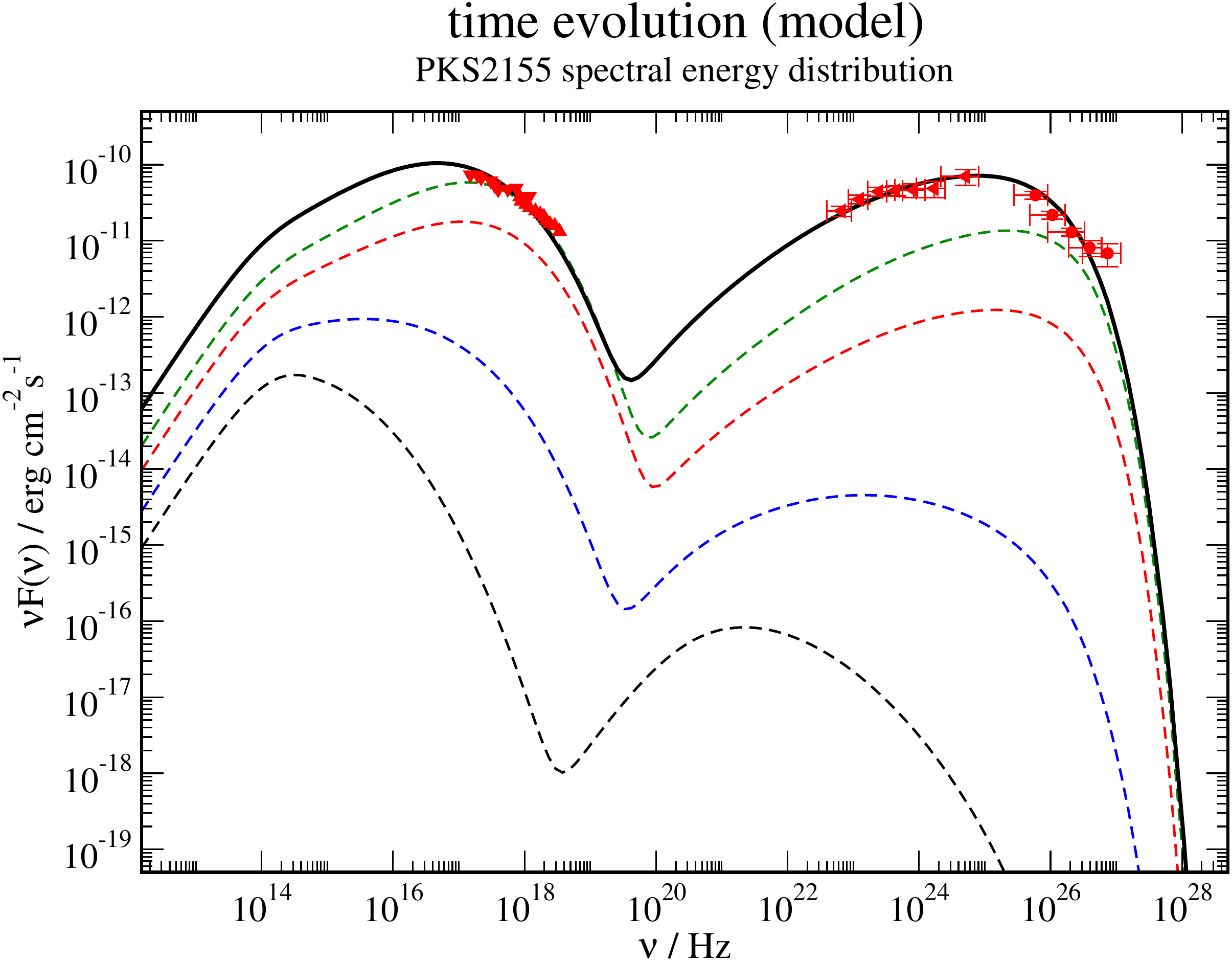}
 \end{center}
 \caption{Time evolution due to the switched on injection function until the steady state of the SED of PKS2155 is reached (see also Fig. 4). The intrinsic times are $t=1\cdot 10^4~\text{s}$ (dashed black), $t=2\cdot 10^4~\text{s}$ (dashed blue), $t=5 \cdot 10^4~\text{s}$ (dashed red), $t=1\cdot 10^5~\text{s}$ (dashed green), the complete steady state is reached at about $t=2\cdot 10^6~\text{s}$, which correlates to the response time of the system due to $t_{\text{esc,rad}}$.}
\end{figure}
\\
The time development of the SED due to a switched on injection of electrons into the acceleration zone at time $t_0 = 0$ is shown in Fig. 3. It can clearly be seen that the final state of the model SSC correlates with the response time of the radiation zone $t_{\text{esc,rad}}$ and that the convergence again begins fast and slows down rapidly at higher simulation times.
\conclusions[CONCLUSIONS]
Our model is able to explain the injection function of many onezone-SSC models as shock and stochastic acceleration  of electrons upstream the jet entering the blob while continuously suffering synchrotron losses. By introducing Fermi-II acceleration we get rid of the sharp cut-off introduced in \citet{kardashev62} or \citet{kirk98} which probably does not occur in physical sources. Additionally we are able to model relatively complex electron densities with increasing and decreasing parts through the stochastic acceleration of electrons, only by varying the monoenergetic injection to higher $\gamma_0$. In contrast to the ad-hoc injection of some onezone-SSC models such Lorentz factors have a physically reasonable, but highly speculative, explanation as upstream previously accelerated but already partially cooled electrons. These electrons are averagely moving in the opposite direction of the blob with a mean Lorentz factor of $\gamma_u \approx 8$ resulting, together with the motion of the blob, in $\gamma_0 \approx 900$ for the monoenergetic injection function used in the acceleration zone of our model.\\ \\
As recent data points out, these complex electron distributions are necessary to model the new constraints concerning the gamma-ray peak of blazar's SEDs if one does not simply shift the synchrotron-peak to achieve the inverse Compton spectrum \citep[e.g.,][]{kataoka00}. The curvature of the peak, and thus the deep dip between the two humps, is a direct consequence of the rising part in the responsible electron distribution within the blob. This constraint rules out many SSC models, which are not able to produce such electron spectra.\\
With our model we are able to form the curvature of the gamma-ray peak and the dip by varying the influence of the Fermi-II processes. The shape and position of the synchrotron peak in the model SED is dominated by $t_{\text{acc}}$ and $R_{\text{acc}}, R_{\text{rad}}$ as well as $B$. For the parameters concerning the acceleration arise from plasmaphysics considerations we gain insight into the jets microphysics while modelling observed SEDs. We have also shown that in such environments the Thomson approximation for the inverse Compton effect can not always be applied, especially when considering time resolution and hence non equilibria of the energy distribution in the blob.
\\ \\
Here we only introduced steady state solutions of our model, but due to the spatially relatively small acceleration region, which is at least an oder of magnitude smaller than the emitting region, this twozone-SSC model is able to selfconsistently model the rising part in the lightcurves of flaring blazars which are connected to the behavior in the acceleration zone, especially the energy transport from low to high energies. This, together with the consequences of the model geometry on the observable SEDs and lightcurves of blazars like in \citet{sokolov}, will be subject of a following paper.

\begin{figure*}[t]
 \vspace*{2mm}
 \begin{center}
 \includegraphics[width=14cm]{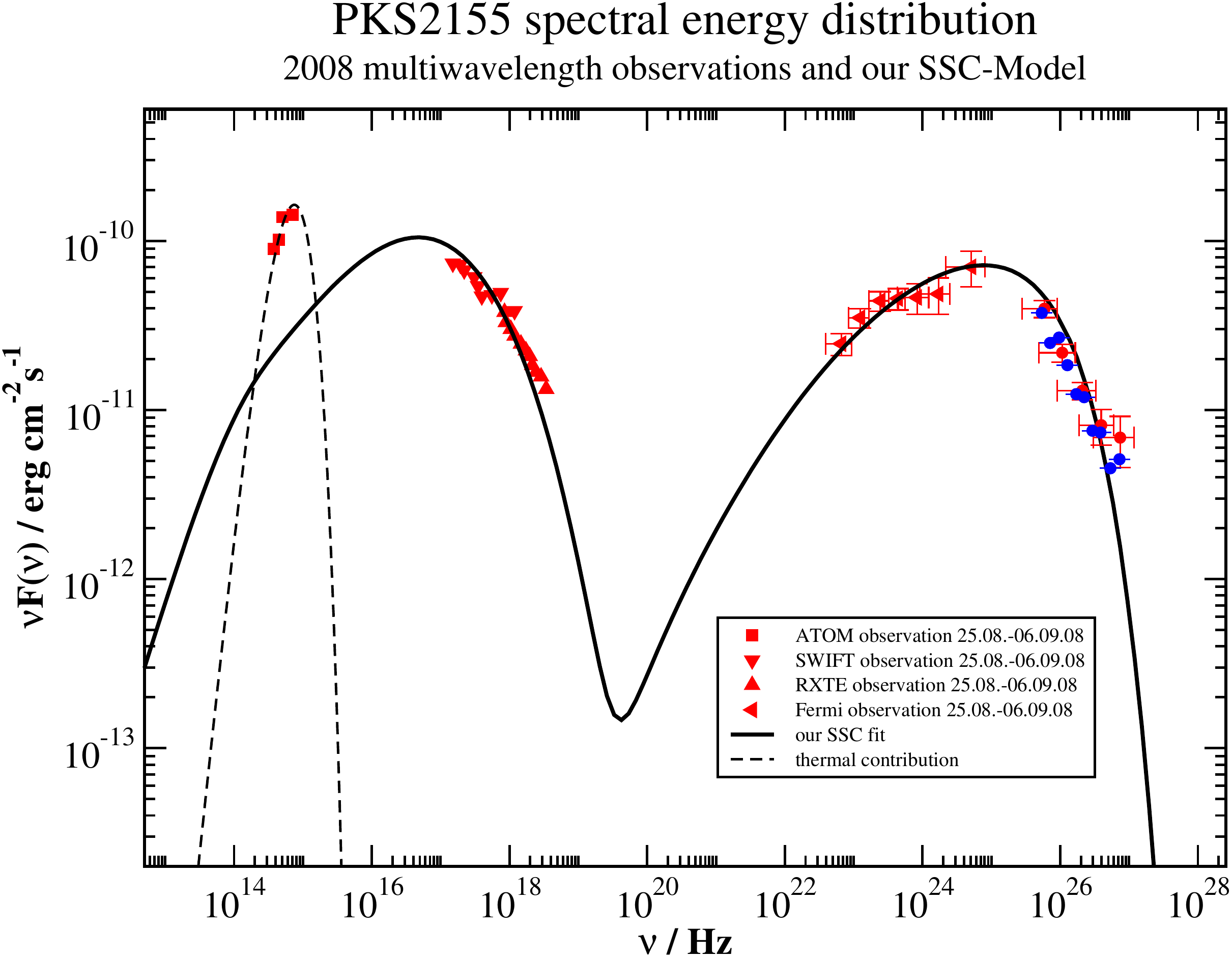}
 \end{center}
 \caption{Lowstate of PKS 2155-30.4 with the simultaneous data of ATOM, SWIFT, RXTE, Fermi and H.E.S.S of the August/September 2008 campaign from \citet{fermi09} (red triangles and circles). The 2003 H.E.S.S. data (blue circles) is also shown, proofing the lowstate of PKS2155-30.4. The VHE data have been deabsorbed using \citet{primack05}. The dashed black curve shows a thermal fit for the contribution of the host-galaxy. Our model SSC fit, arising from the steady state electron distribution in the radiation zone is shown in the solid black curve, a moderate energy injection at $\gamma_0 \approx 910$ into the acceleration zone together with stochastic and systematic acceleration is needed to meet the curvature of the VHE peak given via the Fermi data. }
\end{figure*}


\bibliographystyle{copernicus}
\bibliography{2155_acc}

\addtocounter{figure}{-1}\renewcommand{\thefigure}{\arabic{figure}a}

\end{document}